\UseRawInputEncoding
\documentclass
[superscriptaddress,secnumarabic,amssymb,amsmath,nobibnotes,aps,prd,showkeys,showpacs,nofootinbib,onecolumn,12pt]{revtex4}%
\usepackage{graphicx}
\usepackage{epsf}
\usepackage{bm}
\usepackage{amsmath}
\usepackage{amsfonts}
\usepackage{amssymb}%
\UseRawInputEncoding
\providecommand{\U}[1]{\protect\rule{.1in}{.1in}}

\newcommand{\be}{\begin{equation}}
\newcommand{\ee}{\end{equation}}

\newcommand{\mincir}{\raise
-3.truept\hbox{\rlap{\hbox{$\sim$}}\raise4.truept\hbox{$<$}\ }}
\newcommand{\magcir}{\raise
-3.truept\hbox{\rlap{\hbox{$\sim$}}\raise4.truept\hbox{$>$}\ }}

\begin{document}
\title{Cosmological solutions with time-delay}
\author{Andronikos Paliathanasis}
\email{anpaliat@phys.uoa.gr}
\affiliation{Institute of Systems Science, Durban University of Technology, Durban 4000,
South Africa }
\affiliation{Instituto de Ciencias F\'{\i}sicas y Matem\'{a}ticas, Universidad Austral de
Chile, Valdivia 5090000, Chile}

\begin{abstract}
We introduce a time-delay function in bulk viscosity cosmology. Even for bulk
viscosity functions where closed-form solutions are known, because of the
time-delay term the exact solutions are lost. Therefore in order to study the
cosmological evolution of the resulting models we perform a detail analysis of
the stability of the critical points, which describe de Sitter solutions, by
using Lindstedt's method. We find that for the stability of the critical
points it depends also on the time-delay parameter, where a critical
time-delay value is found which play the role of a bifurcation point. For
time-delay values near to the critical value, the cosmological evolution has a
periodic evolution, this oscillating behaviour is because of the time-delay
function. We find a new behaviour near the exponential expansion point, which
can be seen also as an alternative way to exit the exponential inflation.

\end{abstract}
\date{\today}
\maketitle

\section{Introduction}

\begin{center}
\textit{I will discuss time-delay bulk viscosity cosmology.
Time-delay in cosmological dynamics for the construction of periodic solutions
was one of the last issues examined by the late John D. Barrow. This paper is
dedicated to his memory. }

\bigskip
\end{center}

The need to have a theoretically description of\ the cosmological observations
\cite{dataacc1,dataacc2,data1,data3,data4}, has motivated cosmologists the
last decades to propose different gravitational models, for a review see
\cite{cl1} and references therein. Models with a running vacuum and a particle
production have drown the attention of the scientific society. The main idea
behind these models is that there is production of particles as the universe
is expanded because of the interaction between the gravitational field of the
expanding universe and the quantum vacuum. In the case of a
Friedmann--Lema\^{\i}tre--Robertson--Walker universe the particle production
term is described by a bulk viscosity term for the cosmological fluid in the
gravitational field equations \cite{Sch1939,aniso,aniso2,aniso3}. In the
running vacuum cosmological models the cosmological\ $\Lambda\left(  t\right)
$ term is introduced in the field equations which is an arbitrary function of
time and provide a particle-like production mechanism for the cosmological
fluid \cite{ba1,ba2,ba3,ba4}.

In a isotropic and homogeneous universe, bulk viscosity term is induced by a
divergence of the velocity field for the cosmological fluid which introduce a
effective pressure term for the cosmic fluid. Some first analysis of the bulk
viscosity cosmology based on Eckart's theory \cite{eck1} can be found in
\cite{buv01,buv02}. Bulk viscosity cosmology covers applications for the early
universe and the late-time universe as well. Moreover, other cosmological
models such as Chaplygin gas-like models \cite{ba01,jdbchg,jacobigas,jdan} are
included in the \ bulk viscosity.

Eckart's theory is recovered by the first approximation of the causal theory
of bulk viscosity described by\ Israel and Stewart formalism \cite{isr}.
{Israel-Stewart theory introduce additional degrees of freedom in the field
equations and consequently, it can provide a better explanation on the
physical variables of the observable universe and solves different problems of
Eckart's description, see the discussion in \cite{isr0,isr1}}. Recently a new
approach on the cosmological bulk viscosity proposed in \cite{newap}, the
interesting cosmological applications of that model is that reduces to zero
viscosity in the case of vacuum or a stiff fluid, while for an ideal gas
future singularities can be produced, for more applications of the latter
approach we refer the reader in \cite{newap1}.

{In this work we focus on the most simple bulk viscosity scenario described by
Eckart's formulation,} where we assume that the viscosity term introduce a
that time-delay function in the field equations. The resulting field
equations\ in terms of the Hubble function for a system of first-order delay
differential equations. Time-delay functions play a significant role in
control theory and they are related with the finite time delay response of a
stimulated by the action of an inducer. Applications of time-delay systems
cover all areas of applied mathematics with emphasis in biological systems,
for a review see \cite{tmd}. An interesting discussion on the application of
delay differential equation equations in relativistic physics is given in
\cite{dd1}.

The field equations are reduced in one nonlinear first-order time-delay
differential equation. Exact solutions for time-delay differential equations
are known only for linear time-delay equations. Therefore, we study the
evolution of the nonlinear system near the critical points where for Eckart's
theory the critical points of the field equations describe de Sitter
universes. We find that the evolution of the non time-delay system is
recovered, while when the time-delay is greater or lower from a critical delay
value then we have a new approach to de Sitter solution which is described by
oscillations around the exponential expansion. The critical delay value is the
bifurcation point of the dynamical system, while from time-delay near to the
critical value the oscillations reach an amplitude which can be seen as
constant for large time. Moreover, the stability of the de Sitter solutions is
different in the presence of the time-delay function from the classical case,
and the oscillating behaviour around a unstable de Sitter point can be seen as
a new way to exit the inflationary era in the early universe. Cyclic
cosmological solutions has been widely studied before, however previous
studies are related with cyclic universes around a static Einstein universe
\cite{ccl1,cl2,cl3,cl4,cl5,ccl6}. See also the application of the averaging
approach for the determination of periodic behaviours \cite{gl1,gl2,gl3}. In
\cite{ccl1}, it has been proposed a cyclic cosmological model where the scale
factor has an exponential growth in every cyclic. It was found that such
approach solves various problems of the early universe such as the horizon,
isotropy and flatness problems. The plan of the paper is as follows.

In Section \ref{section2} we give the basic properties and definitions of the
bulk viscosity cosmology. The time-delay bulk viscosity term is introduced in
Section \ref{section3} where Lindstedt's method is applied \cite{com1} in
order to study the stability and the evolution of the dynamical system near
the critical solutions. For the time-delay functions of our consideration
there are two families of critical points which describe the empty space, and
the de Sitter universe. We find that the time-delay function\ affects only the
evolution and the stability of the de Sitter point. In Section \ref{section4},
we repeat our analysis by using dimensionless variables for the dynamical
system from where are able to consider a more general bulk viscosity function.
Finally in Section \ref{section5}, we discuss our results and we extend our
discussion by comparing our results with the simplest formulation of
Israel-Stewart viscosity approach.

\section{Bulk viscosity cosmology}

\label{section2}

In large scales the universe is described by a spatially flat
Friedmann--Lema\^{\i}tre--Robertson--Walker (FLRW) spacetime with line element%
\begin{equation}
ds^{2}=-dt^{2}+a^{2}\left(  t\right)  \left(  dx^{2}+dy^{2}+dz^{2}\right)  ,
\label{bv.01}%
\end{equation}
where for the cosmological fluid source we assume the energy momentum tensor%
\begin{equation}
T_{\mu\nu}=\rho u_{\mu}u_{\nu}+\left(  p+\eta\right)  h_{\mu\nu} \label{bv.02}%
\end{equation}
in which $u^{\mu}=\delta_{t}^{\mu}$ is the comoving observer and $h_{\mu\nu
}=g_{\mu\nu}+u_{\mu}u_{\nu}$ is the projective tensor. Functions $\rho$ and
$p$ describe the energy density and pressure of the perfect fluid while
$\eta=\eta\left(  \rho\right)  $ is the bulk viscosity term
\cite{bv01,bv02,bv03,bv04,bv05,bv06}.

The introduction of the bulk viscosity term modify the Einstein field
equations as follows%
\begin{align}
3H^{2}  &  =\rho~\label{bv.03}\\
2\dot{H}+3H^{2}  &  =-p_{eff} \label{bv.04}%
\end{align}
where now $p_{eff}$ is the effective pressure term given by the expression
$p_{eff}=p-\eta\left(  t\right)  $.

The bulk viscosity term can take various functional forms such that to
describe other cosmological models as the Chaplygin gas and its modifications
\cite{akam,bento,zhi,ch1,ch2}. Chaplygin gas model is recovered when the
perfect fluid is an ideal gas, i.e. $p=\left(  \gamma-1\right)  \rho$ and
$\eta=\eta_{0}\rho^{-\lambda}$. The later model has been proposed a unified
dark matter theory, while it is possible to be applied in the early universe
\cite{amen1}, and specifically in the description of inflation
\cite{ba01,jdbchg,jacobigas,jdan}.

The conservation law $T_{~~;\nu}^{\mu\nu}=0$ has the following nonzero
component%
\begin{equation}
\dot{\rho}+3H\left(  \rho+p\right)  =3H\eta. \label{bv.05}%
\end{equation}
where we can see that\ the rhs of the conservation law introduce a particle
creation/destroy term. \ Particle production process might be considered as
another approach to explain different phases of the universe's evolution and
it has various applications in the early and late universe also
\cite{b01,b02,b03,b04,b05,b06,b07,b08,b09}.

We assume that the perfect fluid is an ideal gas and $\eta\left(  t\right)
=\eta\left(  H\right)  $, then from (\ref{bv.03}) and (\ref{bv.04}) we end
with the following first-order ordinary differential equation%
\begin{equation}
2\dot{H}+3\gamma H^{2}-\eta\left(  H\right)  =0. \label{bv.06}%
\end{equation}

Equation (\ref{bv.06}) can be solved explicitly by quadratures, indeed the
generic solution is
\begin{equation}
\int\frac{2dH}{\eta\left(  H\right)  -3\gamma H^{2}}=t-t_{0} \label{bv.07}%
\end{equation}

For specific forms of the bulk viscosity the differential equation
(\ref{bv.06}) takes the form of some known equations form where we can write
the analytic solution in a closed-form expression. For instance when
$\eta\left(  H\right)  $ is linear, then (\ref{bv.06}) takes the form of the
Riccati first-order ODE, while when $\eta\left(  H\right)  $ is a polynomial
function of order three, equation (\ref{bv.06}) is the Abel equation.

Let $H_{0}\in%
\mathbb{R}
$ be a zero of the function $f\left(  H_{0}\right)  =0$, where%
\begin{equation}
f\left(  H\right)  =\frac{\eta\left(  H\right)  -3\gamma H^{2}}{2}.
\label{bv.08}%
\end{equation}
From a physical point of view, for $H_{0}\neq0$, the critical point describe a
de Sitter universe while for $H_{0}=0$ the resulting spacetime is the empty
Minkowski space. The critical point of the differential equation (\ref{bv.06})
will be an attractor when $\frac{df}{dH}|_{H=H_{0}}<0$, or a saddle point when
$\frac{df}{dH}|_{H=H_{0}}>0$. Since function $\eta\left(  H\right)  $ is a
real function then there is not any possibility around the critical point to
have a periodic behaviour. The question that arise is if it is possible to get
a periodic behaviour for the first-order differential equation without
assuming any complex function $\eta\left(  H\right)  .$ {At this point it is
important to mention that we refer to a periodic behaviour for the Hubble
function and not for the scalar factor, thus the periodic behaviour does not
refer necessary to a bounce cosmology, but in general to a cyclic universe. }

In other natural sciences, such as biology, epidemiology or either in signal
analysis, a mathematical approach to achieve such periodic behaviour around is
thought the introduction of a time-delay term. Time delays reflects the time
taken to detect the error signal and the time taken to act on it \cite{tmd}.
Applications of time-delay can be found also in astronomy and astrophysics,
for instance see in \cite{td01,td03} and references therein.

We demonstrate how the introduction of a time-delay can introduce a periodic
behaviour by considering the first-order linear delay ordinary differential
equation%
\begin{equation}
\dot{x}\left(  t\right)  +\alpha x\left(  t-T\right)  =0. \label{bv.09}%
\end{equation}
In the special case where~$a=1,$ $T=\frac{\pi}{2}$ then we can see that a
solution of (\ref{bv.09}) is $x\left(  t\right)  =x_{0}\cos\left(  t\right)  $.

The generic solution of equation (\ref{bv.09}) is given by $x\left(  t\right)
=x_{0}e^{\lambda t}$ where $\lambda$ is a solution of the characteristic
equation
\begin{equation}
\lambda+\alpha e^{-\lambda T}=0. \label{bv.11}%
\end{equation}
The characteristic equation can provide complex values for $\lambda$ \ when
$\alpha T>0$, which means that $x\left(  t\right)  $ can be a periodic function.

\section{Bulk viscosity $\eta\left(  H\right)  $ with time-delay}

\label{section3}

We make use of the free-function which describes the viscosity in order to
introduce the time-delay in the cosmological studies. The delay parameter
measures the time-difference between the compression or expansion of the fluid
and the necessary time in order the equilibrium to be restored{.}

\subsection{Linear bulk viscosity}

Consider the simplest time-delay bulk viscosity function $\eta\left(
H\right)  =2\eta_{0}H\left(  t-T\right)  $ which introduces a delay term on
the proposed model in \cite{packer1}. Equation\ (\ref{bv.06}) becomes
\begin{equation}
2\dot{H}\left(  t\right)  +3\gamma\left(  H\left(  t\right)  \right)
^{2}-2\eta_{0}H\left(  t-T\right)  =0. \label{bv.12}%
\end{equation}
Equation (\ref{bv.12}) admits two critical points, they are $H_{A}=0$ and
$H_{B}=\frac{2\eta_{0}}{3\gamma}$. \

Point $H_{A}$ describes an empty universe, while $H_{B}$ denotes a de Sitter
phase, which correspond to the de Sitter inflation \cite{in1,in2}.

At the critical point $H_{A}$, the linearized equation is written as
\begin{equation}
\dot{H}\left(  t\right)  -\eta_{0}H\left(  t-T\right)  =0. \label{bv.13}%
\end{equation}
Hence, according to the discussion we did before there is a periodic behaviour
around $H_{A}$ iff $\eta_{0}T<0$.

We perform the change of variable $H\left(  t\right)  =y\left(  t\right)
+\frac{2\eta_{0}}{3\gamma}$, such that $H_{B}$ corresponds to $y_{B}=0$. In
the new variable, equation (\ref{bv.12}) is written%
\begin{equation}
\dot{y}+\frac{3}{2}\gamma y^{2}+2\eta_{0}y-\eta_{0}y\left(  t-T\right)  =0,
\label{bv.14}%
\end{equation}
where near to the critical point $y_{B}$ \ the later equation is linearized as
follows
\begin{equation}
\dot{y}+2\eta_{0}y-\eta_{0}y\left(  t-T\right)  =0. \label{bv.15}%
\end{equation}

Equation (\ref{bv.15}) admits an oscillating solutions when the parameters
$\eta_{0}$ and $T$ satisfy the condition \cite{gyoribook}
\begin{equation}
-\eta_{0}Te^{2\eta_{0}T}>e^{-1}. \label{bv.16}%
\end{equation}
We infer that the later expression can not be true for any $\eta_{0},T$, that
is, there are not oscillations around the critical point $H_{B}$.

We demonstrate it by replacing $y\left(  t\right)  =A\cos\left(  \omega
t\right)  $ in (\ref{bv.15}) from where find the algebraic equations
\begin{equation}
2-\cos\left(  \omega T\right)  =0~,~\omega+\eta_{0}\sin\left(  \omega
T\right)  =0
\end{equation}
with solution%
\begin{equation}
\omega^{2}=-\sqrt{3}\text{ and }T^{2}=-\frac{\arccos2}{\sqrt{3}}%
\end{equation}
Because $\omega$ is a pure imaginary number we conclude that there is not any
oscillating behaviour around the critical point $H_{B}$.

The stability of the critical point $H_{B}$ can be determined easily from the
linearized equation (\ref{bv.15}). Equation admits the exact solution
$y\left(  t\right)  =y_{0}e^{\lambda t}$ where $\lambda$ is a solution of the
algebraic equation
\begin{equation}
\left(  2-e^{-\lambda T}\right)  \eta_{0}+\lambda=0.
\end{equation}
From the last expression we find that the point $H_{B}$ is a source when
$\lambda<0$, that is $\eta_{0}T>0,~T>0$ , or $\eta_{0}T<0,~T<0$. We can see
that there are differences in the stability when the time-delay is positive or negative.

Recall that for that specific model with linear bulk viscosity for the
nondelay equation point $H_{A}$ is an attractor when $\eta_{0}<0$, while point
$H_{B}$ is an attractor when $\eta_{0}>0$.

In equation (\ref{bv.12}) we perform the scale transformation~$t=-\frac{\tau
}{\omega\eta_{0}}$ and the rescale $H=\varepsilon h$ such that equation
(\ref{bv.12}) to be written as follows%
\begin{equation}
\omega\frac{dh}{d\tau}-\frac{3\gamma}{2\eta_{0}}\varepsilon\left(  h\left(
\tau\right)  \right)  ^{2}+h\left(  \tau+\omega\eta_{0}T\right)  =0
\label{bv.17}%
\end{equation}
where $\varepsilon<<1$ in order $H$ to take values near the critical point
$H_{A}$.

Assume now that $-\eta_{0}T=\frac{\pi}{2}+\varepsilon^{2}\mu$, and$~\omega
=1+\varepsilon\omega_{1}+...$ then by performing a Taylor expansion of
$h=h_{0}+\varepsilon h_{1}+...$ we get%
\begin{align}
h_{0}^{\prime}+h_{0}\left(  \tau-\frac{\pi}{2}\right)   &  =0\\
h_{1}^{\prime}+h_{1}\left(  \tau-\frac{\pi}{2}\right)  +\frac{3\gamma}%
{2\eta_{0}}h_{0}^{2}  &  =0\\
h_{2}^{^{\prime}}+h_{2}\left(  \tau-\frac{\pi}{2}\right)  +\omega_{2}%
h_{0}^{\prime}-\left(  \frac{\pi}{2}\omega_{2}+\mu\right)  h_{0}\left(
\tau-\frac{\pi}{2}\right)  -\frac{3\gamma}{\eta_{0}}h_{0}h_{1}  &  =0\\
&  ...
\end{align}

From the two first equations we get
\begin{align}
h_{0}\left(  \tau\right)   &  =A\cos\tau~,\label{bv.18}\\
h_{1}\left(  \tau\right)   &  =\frac{3\gamma}{4\eta_{0}}A^{2}\left(
1+\cos\left(  2\tau\right)  \right)  \label{bv.19}%
\end{align}
and by replacing in the differential equation for $h_{2}\left(  \tau\right)  $
and require the coefficients of $\cos\tau$ and $\sin\tau$ to be zero we get%
\begin{equation}
\omega_{2}=0~,~A^{2}=-\left(  \frac{2\eta_{0}}{3\gamma}\right)  ^{2}\mu
\end{equation}

From where we infer that the critical point is stable when $\mu<0.$ That means
that for$~-\eta_{0}T<\frac{\pi}{2}$ the critical point is a source. The
amplitude in the original coordinates is given by the expression%
\begin{equation}
A_{H}^{2}=\left(  \frac{2\eta_{0}}{3\gamma}\right)  ^{2}\left(  \frac{\pi}%
{2}-\eta_{0}T\right)  .
\end{equation}

\begin{figure}[t]
\includegraphics[scale=0.65]{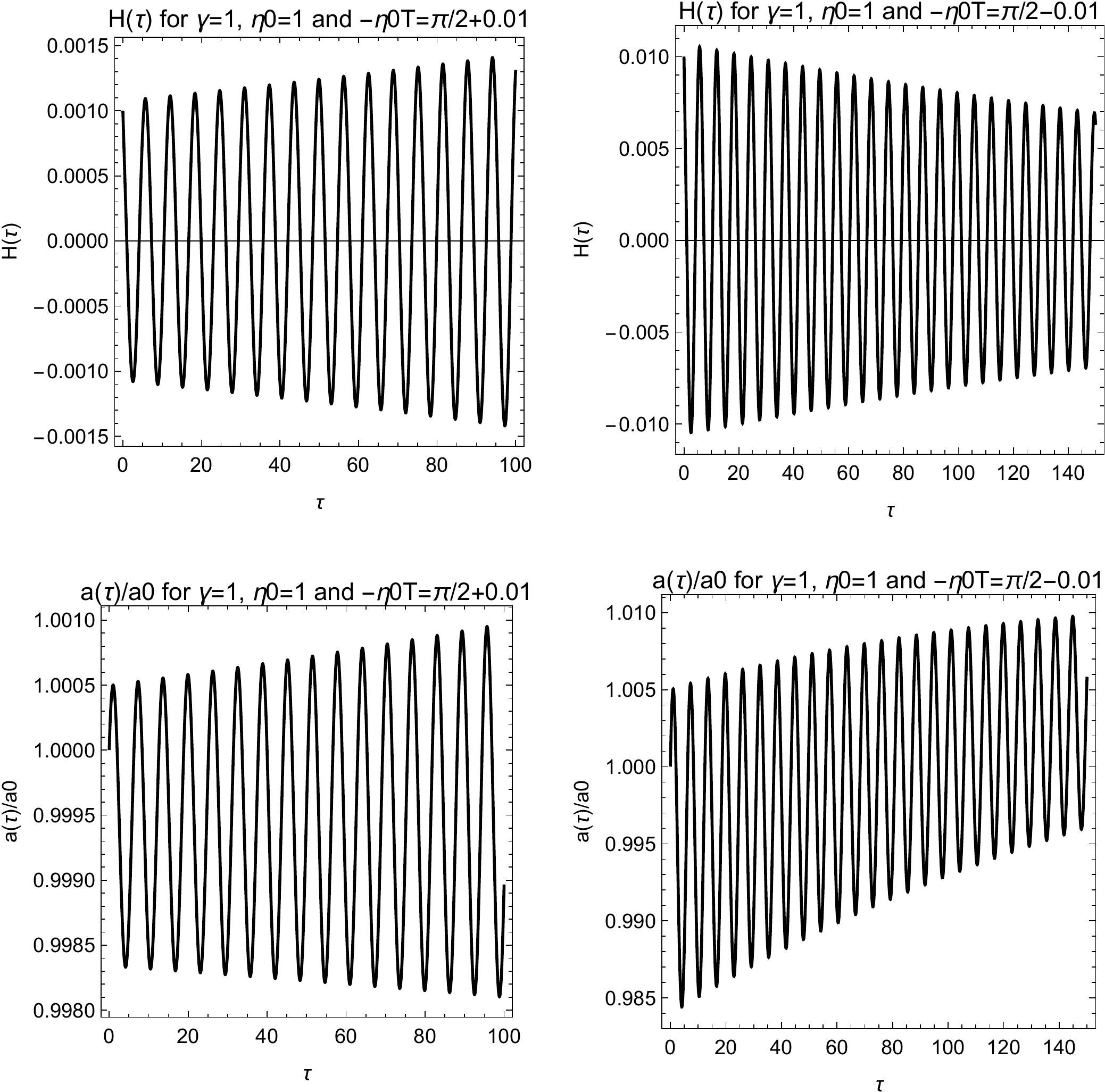} \caption{Numerical simulation of
equation (\ref{bv.12}) around the critical point $H_{A}=0$. The simulation is
for $\gamma=1$ and $\eta_{0}=1.$ Upper figures the evolution of $\ H\left(
\tau\right)  $ is presented, while in the lower figures the scale factor
$\frac{a\left(  \tau\right)  }{a_{0}}=e^{\int H\left(  \tau\right)  d\tau}$ is
given. Left Figs. are for delay $-\eta_{0}T=\frac{\pi}{2}+0.01$, and the
critical point is unstable while right Figs. are for $-\eta_{0}T=\frac{\pi}%
{2}-0.01$ where the critical point is an attractor. The value $-\eta
_{0}T=\frac{\pi}{2}$ is a critical delay and a Hopf bifurcation point.}%
\label{fig1}%
\end{figure}

\begin{figure}[t]
\includegraphics[scale=0.65]{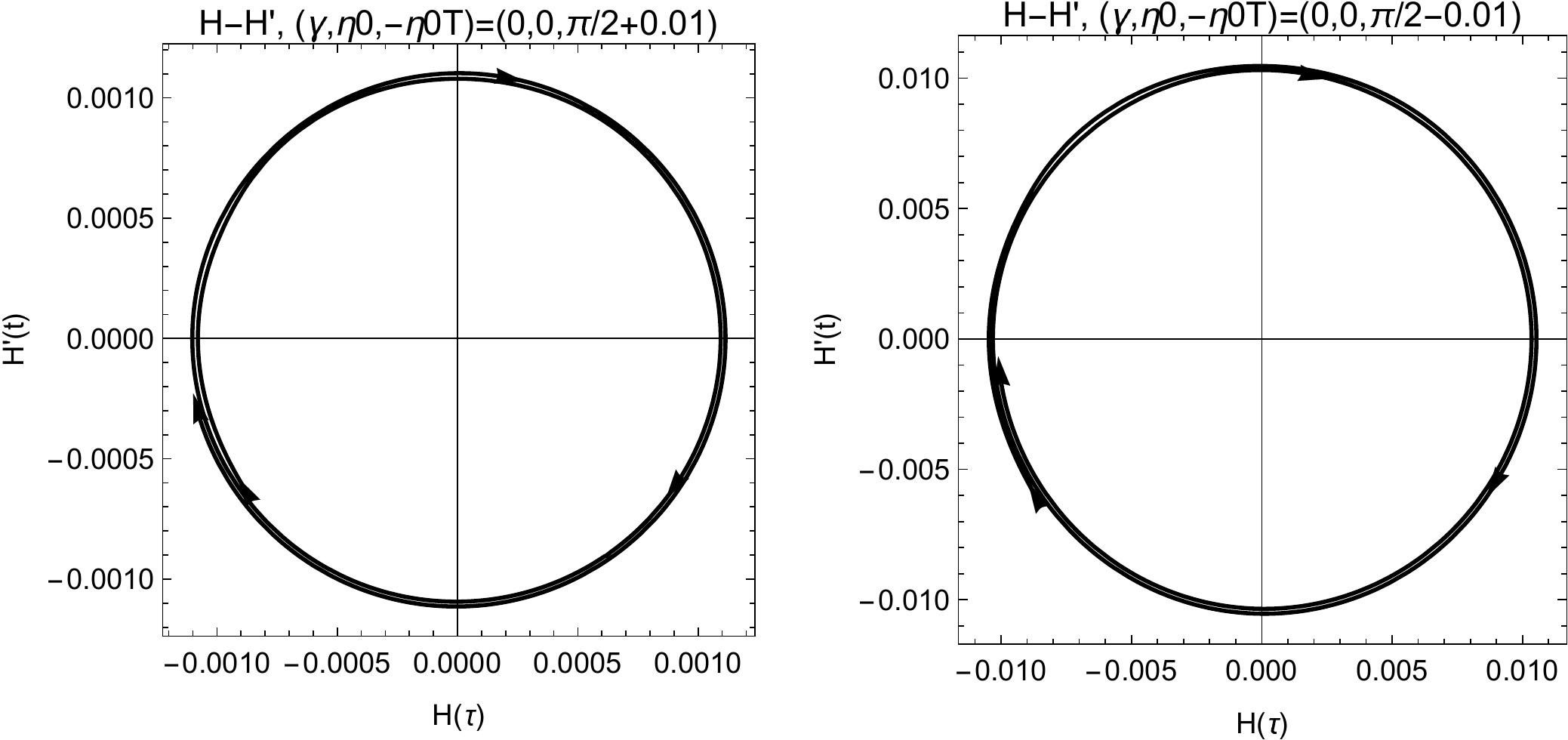} \caption{Phase-space diagram for the
solution of equation (\ref{bv.12}) near the critical point $H_{A}=0$. The
simulation is for $\gamma=1$ and $\eta_{0}=1.$ Left Fig. are for delay
$-\eta_{0}T=\frac{\pi}{2}+0.01$, and the critical point is unstable while
right Fig. is for $-\eta_{0}T=\frac{\pi}{2}-0.01$ where the critical point is
an attractor. }%
\label{fig2}%
\end{figure}

The stability of the critical point is demonstrate in Fig. \ref{fig1} where
the evolution of $H\left(  \tau\right)  $ is given around the critical point.
We observe that for $-\eta_{0}T<\frac{\pi}{2}$ the solution oscillates around
the Minkowski solution, while in the second case where $\,-\eta_{0}T>\frac
{\pi}{2}$ the critical point is unstable. In \ref{fig2} the phase-space
diagram for the solution near the critical point $H_{A}$ is presented.

With the use of (\ref{bv.18}) and (\ref{bv.19}) the scale factor around the
critical point is given by the following expression
\begin{equation}
\ln\left(  \frac{a\left(  \tau\right)  }{a_{0}}\right)  =\frac{\eta_{0}%
}{6\gamma}\left(  4\sqrt{\left(  \frac{\pi}{2}-\eta_{0}T\right)  }\sin
\tau+\varepsilon\left(  \frac{\pi}{2}-\eta_{0}T\right)  \left(  2\tau
+\sin\left(  2\tau\right)  \right)  \right)  +O\left(  \varepsilon^{2}\right)
\end{equation}
from where it is clear that for $-\eta_{0}T<\frac{\pi}{2}$ the scale factor
oscillates around the constant solution.

\subsection{Bulk viscosity $\eta\left(  H\right)  =-3\eta_{1}H^{2}\left(
t-T\right)  +2\eta_{0}H\left(  t-T\right)  $}

Consider now the bulk viscosity to be $\eta\left(  H\right)  =-3\eta_{1}%
H^{2}\left(  t-T\right)  +2\eta_{0}H\left(  t-T\right)  ,~\gamma+\eta_{1}%
\neq0,$ where now equation\ (\ref{bv.06}) becomes%
\begin{equation}
2\dot{H}\left(  t\right)  +3\gamma\left(  H\left(  t\right)  \right)
^{2}+3\eta_{1}H^{2}\left(  t-T\right)  -2\eta_{0}H\left(  t-T\right)  =0.
\label{bvv.01}%
\end{equation}

When there is not any delay, i.e. $T=0$, the later dynamical system can been
as a linear bulk viscosity function $\eta\left(  H\right)  $ for an ideal gas
with equation of state parameter $\bar{\gamma}=\gamma+\eta_{1}$. Thus there
are two critical points, the static solution $\bar{H}_{A}=0$ and the de Sitter
point with $\bar{H}_{B}=\frac{2\eta_{0}}{3\left(  \gamma+\eta_{1}\right)  }$.

For the static solution, i.e. $\bar{H}_{A}$ the stability analysis is that for
the linear bulk viscosity function, hence we omit it.

We focus on the stability analysis for the de Sitter point $\bar{H}_{B}$.$~$We
perform the change of variables $H\left(  t\right)  =\bar{H}_{B}+y\left(
t\right)  $ and equation (\ref{bvv.01}) becomes%
\begin{equation}
2y^{\prime}+\frac{4\eta_{0}\gamma}{\eta_{1}+\gamma}y+2\eta_{0}\left(
\frac{\eta_{1}-\gamma}{\eta_{1}+\gamma}\right)  y\left(  t-T\right)  +3\gamma
y^{2}+3\eta_{1}y^{2}\left(  t-T\right)  =0. \label{bvv.02}%
\end{equation}

Near to the critical point $y_{B}=0$, for linearized system we replace
$y\left(  t\right)  =A\cos\left(  \omega t\right)  $ from where we find%
\begin{equation}
\cos\left(  \omega T_{cr}\right)  =-\frac{2\gamma}{\eta_{1}-\gamma}%
~,~\sin\left(  \omega T_{cr}\right)  =\frac{\left(  1+\eta_{1}\right)  }%
{\eta_{0}\left(  \eta_{1}-1\right)  }%
\end{equation}
where we find%
\begin{equation}
\omega^{2}=\left(  \eta_{1}-3\gamma\right)  \left(  \eta_{1}+\gamma\right)
\left(  \frac{\eta_{0}\left(  \eta_{1}-1\right)  }{\left(  \gamma-\eta
_{1}\right)  ^{2}\left(  \eta_{1}+1\right)  }\right)  ^{2},
\end{equation}%
\[
T_{cr}=\frac{\left(  \gamma-\eta_{1}\right)  \left(  \eta_{1}+1\right)
}{2\gamma\left(  \eta_{1}-1\right)  }.
\]
Therefore, at this case it is possible to have a periodic behaviour the de
Sitter solution assuming that the following relation holds%
\begin{equation}
\left(  \eta_{1}-3\gamma\right)  \left(  \eta_{1}+\gamma\right)  >0.
\end{equation}

In order to study the stability of the solution we work as before and we do
the change\ of variables $\tau=\Omega t$ such that equation (\ref{bvv.02})
becomes%
\begin{equation}
2\Omega y^{\prime}+\frac{4\eta_{0}\gamma}{\eta_{1}+\gamma}y+2\eta_{0}\left(
\frac{\eta_{1}-\gamma}{\eta_{1}+\gamma}\right)  y\left(  t-\Omega T\right)
+3\gamma y^{2}+3\eta_{1}y^{2}\left(  t-\Omega T\right)  =0.
\end{equation}

Similarly with above we replace $\Omega=\omega+\varepsilon^{2}k_{2}%
+...,~y=\varepsilon y_{0}+\varepsilon^{2}y_{1}+...$ and $T=T_{cr}%
+\varepsilon^{2}\mu+...$ in the later equation where we find that the
amplitude $A$ of the oscillations is expressed as follows%
\begin{equation}
A^{2}=\frac{8\eta_{0}^{4}\left(  \eta_{1}-\gamma\right)  \omega^{2}}{9P\left(
A\right)  }\mu
\end{equation}
and
\begin{align}
P\left(  A\right)   &  =\left(  3\gamma^{3}-\gamma\eta_{1}+2\eta_{1}%
^{3}\right)  \left(  \gamma+2\eta_{0}T_{cr}+\eta_{1}\right)  \gamma\eta
_{0}^{3}+\eta_{1}^{2}T_{cr}\left(  \gamma+\eta_{1}\right)  ^{3}\omega
^{4}\nonumber\\
&  +\eta_{0}\eta_{1}\left(  \gamma+\eta_{1}\right)  \left(  \gamma^{3}\left(
3\eta_{0}T_{cr}-1\right)  -3\gamma\eta_{1}^{2}\left(  \eta_{0}T_{cr}-1\right)
+2\eta_{0}\eta_{1}^{3}T_{cr}+2\eta_{1}\gamma^{2}\left(  3\eta_{0}%
T_{cr}+1\right)  \right)  \omega^{2}.
\end{align}
Recall that for $A^{2}>0$ the critical point is an attractor, while when
$A^{2}<0$, the critical point is a source.

\section{Evolution in dimensionless variables}

\label{section4}

We continue our analysis by considering the dimensionless variables%
\begin{equation}
\Omega=\frac{\rho}{3H^{2}}~,~\Delta=\frac{\eta}{3\rho}. \label{bbv.01}%
\end{equation}

Such variables are useful because the existence of ideal gas solutions,
power-law solutions, can be investigated. For the bulk viscosity function we
assume the case where $\eta\sim\rho\left(  t-T\right)  ^{\nu}$, the field
equations are reduced to the following first order time-delay ordinary
differential equation%
\begin{equation}
\Delta^{\prime}=3\Delta\left[  3\left(  \nu\Delta\left(  N-T\right)
-\Delta\right)  -\left(  \nu-1\right)  \gamma\right]  \label{bbv.02}%
\end{equation}
where $\Delta^{\prime}=\frac{d\Delta}{dN},~N=\ln a,$ while$~$from the first
Friedmann equation $\Omega=1$, and
\begin{equation}
\frac{2\dot{H}}{3H^{2}}=\left(  3\Delta-\gamma\right)  \label{bbv.03}%
\end{equation}

Equation (\ref{bbv.02}) admits two critical points, point $P_{1}$, with
$\Delta=0$ and $P_{2}$ $\ $with $\Delta=\frac{\gamma}{3}$.

The solution at $P_{1}$ describes a universe without the bulk viscosity term,
while point $P_{2}$ a de Sitter universe where the bulk viscosity term
contributes in the evolution of the universe.

As far as the stability of the critical point $P_{1}$ is concerned, we replace
$\Delta=\varepsilon\delta$ in (\ref{bbv.02}) and we linearize around
$\varepsilon=0,$ we find
\begin{equation}
\delta^{\prime}=3\left(  1-\nu\right)  \gamma\delta\left(  N\right)  ,
\label{bbv.04}%
\end{equation}
from where can infer that the point is stable when $\left(  1-\nu\right)
\gamma<0$.

For $P_{2}$ we perform the change of variable~$\Delta=\frac{\gamma}%
{3}+\varepsilon\delta$ with $\varepsilon^{2}=0$, we find%
\begin{equation}
\delta^{\prime}=3\gamma\left(  \delta\left(  N\right)  -\nu\delta\left(
N-T\right)  \right)  . \label{bbv.05}%
\end{equation}

An exact solution of the latter equation is $\delta\left(  t\right)
=\delta_{0}\cos\left(  \omega t\right)  $, where
\begin{equation}
\left(  \frac{\omega}{3\gamma}\right)  ^{2}=\nu^{2}-1~,~\text{and }%
T=T_{cr}=-\frac{1}{\omega}\arccos\frac{1}{\nu}.
\end{equation}
Consequently, there is a periodic behaviour around the de Sitter point if and
only if $\left\vert \nu\right\vert >1$. For $\nu<1$ the time-delay is
positive, while for $\nu>1$ the time-delay parameter is negative in order a
periodic behaviour to exist.

In order to study the stability of the critical point $P_{2}$ we follow the
same steps as above from where we find that the amplitude of the oscillations
is given by the expression
\begin{equation}
\left(  \delta_{0}\right)  ^{2}=-\frac{4\gamma^{3}\mu\nu\left(  4\nu^{4}%
+13\nu^{2}-8\right)  }{\nu\left(  \nu\left(  \nu\left(  8\nu^{2}%
+4\nu+47\right)  +10\right)  -64\right)  -32}, \label{bbv.06}%
\end{equation}
where for $\left(  \delta_{0}\right)  ^{2}>0$ the critical point is stable and
$\mu$ is$~\mu=T-T_{cr}$.

Easily we observe that for $\nu=\frac{1}{2}$ in which $\eta\sim H\left(
N-T\right)  $ there is not a a periodic behaviour around the de Sitter
solution, a result in agreement with the analysis did above for the critical
point $H_{B}$.

Note that the Minkowski spacetime is not recovered, since $\Delta~$is finite.
That can be recovered by doing the change of variables. Since such solution is
not of our interest we omit the analysis.

\begin{figure}[t]
\includegraphics[scale=0.65]{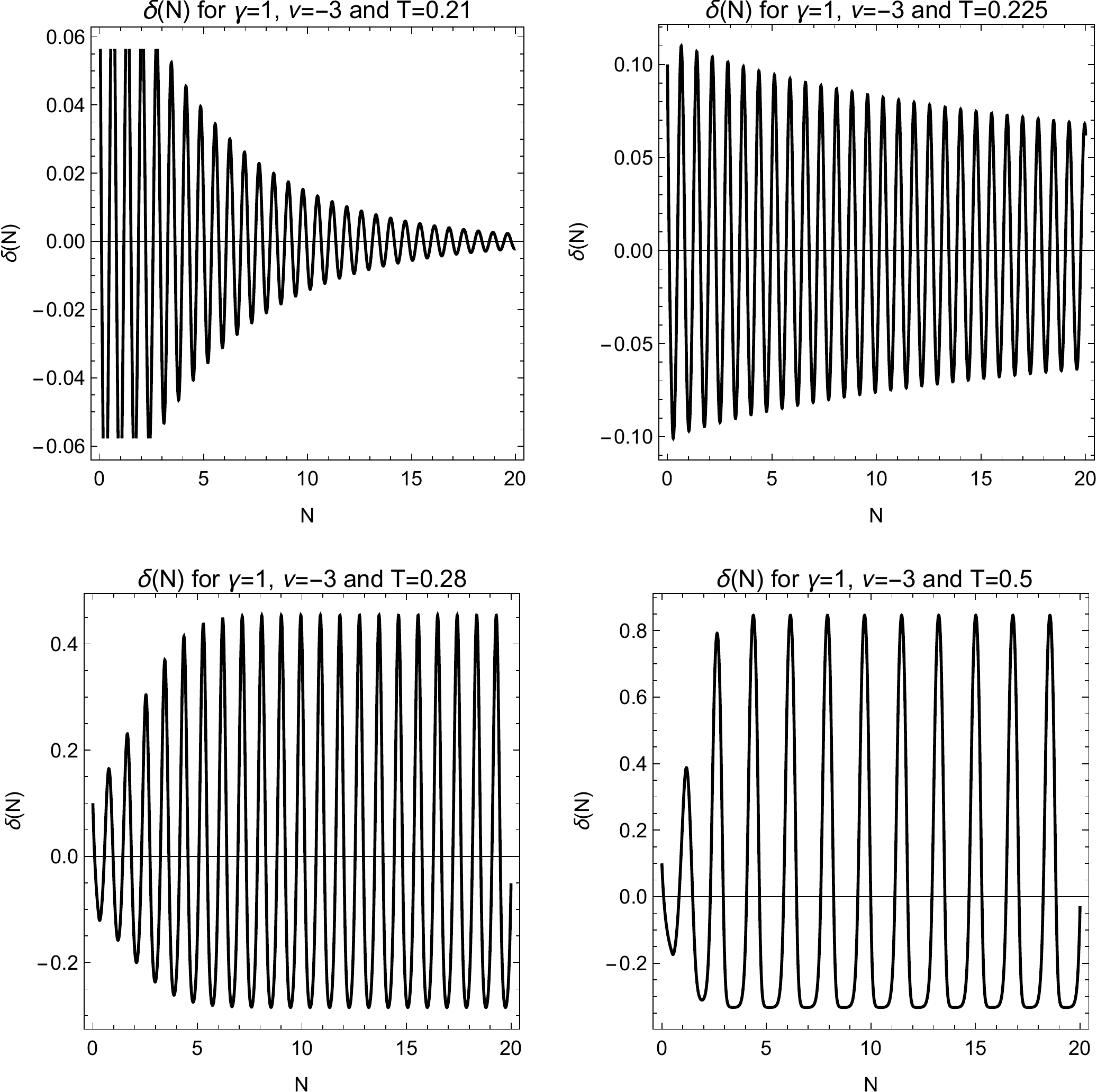} \caption{Numerical simulation of
equation (\ref{bbv.02}) around the de Sitter point $\Delta=\frac{\gamma}%
{3}+\delta\left(  N\right)  $. The simulation is for $\gamma=1$ and $\nu=-3$
where~$T_{cr}=0.23.$ Upper figures are for $T<T_{cr}$ where the critical point
is an attractor while lower figures are for $T>T_{cr}$ and the critical point
is unstable. }%
\label{fig4}%
\end{figure}

In Fig. \ref{fig4} we plot the numerical evolution of equation (\ref{bbv.02})
for $\Delta=\frac{\gamma}{3}+\delta\left(  N\right)  $, for four different
values of the delay parameter. Moreover, by using expression (\ref{bbv.03}) we
can write the equation of state parameter to be $w=-1+\delta\left(  N\right)
,$ from where we observe that there are oscillations around the de Sitter
solution. Which is another way to reach or exit the exponential inflation
point in bulk viscosity theory \cite{in1,in2}.

Function $w\left(  N\right)  $ near to the critical point is approximated as
follows%
\begin{equation}
w\left(  N\right)  =-1+\varepsilon\delta_{0}\cos\left(  \omega N\right)
+O\left(  \varepsilon^{2}\right)  . \label{bbv.07}%
\end{equation}

{Equation of state parameters described by periodic functions studied before
in \cite{mm0} which were compared with the cosmological observation.}

\section{Conclusion}

\label{section5}

In this work we introduced a time-delay in Eckart's formulation of bulk
viscosity cosmology. Time-delay has many applications in the optimal control
theory for the applied sciences, where the delay describe the finite
time-response of actuators which have been used in the implementation of the
control law.

In the presence of the time-delay, the theory remains of second-order in terms
of the scale factor or of first-order in terms of the Hubble function. Without
the time-delay function such cosmological models describe singular solutions
and a de Sitter evolution which correspond to an exponential inflation era.
However, when a small parameter is introduced the evolution of the
cosmological model changes dramatically near the de Sitter solution.
Specifically, because of the time-delay function the Hubble function has an
oscillating behaviour around a constant value which correspond to the de
Sitter universe. Such a behaviour can not be provided by a real bulk viscosity
term in Eckart's theory.

There are other treatments for the bulk viscosity theory, such is the
Israel-Steward formalism, where the bulk viscosity term $\eta\left(  t\right)
$ satisfies the first-order ordinary differential equation \cite{ma1}.
Israel-Steward theory is a second-order theory in terms of the Hubble
function. In\ simplest case of the Israel-Steward theory the bulk viscosity
term satisfies the following equation%
\begin{equation}
\tau\dot{\eta}+\eta=3\xi H \label{bbv.10}%
\end{equation}
where $\tau$ is the relaxation time and it is assumed to be $\tau=\xi\rho
^{-1}$ where $\xi$ is the bulk viscosity coefficient. Equation (\ref{bbv.10})
can equivalently written as follows%
\begin{align}
\dot{H}-Y  &  =0\\
2\dot{Y}+6\gamma HY+3H^{2}\left(  \left(  2Y+3\gamma H^{2}\right)  \xi
^{-1}-H\right)   &  =0.
\end{align}
For \ $\xi=3\gamma\xi_{0}\rho^{\kappa}$, the latter system admits the real
critical point which describe the de Sitter solution $Y=0$, $H_{P}=\xi
_{0}^{\frac{1}{1-\kappa}}$. The matrix of the linearized system near the
critical point admits imaginary eigenvalues when $\kappa<-1-\frac{1}%
{3\gamma^{2}}-3\gamma^{2}$. That means that the point which describe the de
Sitter solution is a spiral. However from phenomenological point of view
parameter $\kappa$ should be positive \cite{bv01}, i.e.$\kappa>0,$ where there
is not any oscillating behaviour. In contrary with the time-delay model
studied before where oscillating behaviour was found also for positive values
of $\kappa.$

In the full formulation of the Israel-Steward theory the bulk viscosity
satisfies the ordinary differential equation%

\begin{equation}
\tau\dot{\eta}+\eta+\frac{1}{2}\left(  3Ç +\frac{\dot{\tau}}{\tau}%
-\frac{\dot{\xi}}{\xi}-\frac{\dot{T}}{T}\right)  \tau\eta=\xi H.
\label{bbv.08}%
\end{equation}
Because of the arbitrariness of the functions $\tau$ and $\xi$, different
cosmological evolutions can be recovered \cite{ma1}. The effects of a
time-delay term in the full Israel-Steward theory is the subject of study of a
future work.

In \cite{bgin} inspired by the bulk viscosity cosmology, a new class of exact
solutions of the field equations determined for a fluid with a equation of
state parameter $\rho\left(  t\right)  +p\left(  t\right)  =\alpha\rho\left(
t\right)  ^{\nu}$. The field equation reduce to the following first-order
differential equation $\dot{H}+\bar{\alpha}H^{2\nu}=0$where $\bar{\alpha
}=2^{-1}3^{\nu}\alpha$. The latter equation has the following solution%
\begin{align}
H\left(  t\right)   &  =\left(  \bar{\alpha}\left(  2\nu-1\right)  \left(
t-t_{0}\right)  \right)  ^{\frac{1}{1-2\nu}}~,~\nu\neq\frac{1}{2},\\
H\left(  t\right)   &  =e^{-\bar{\alpha}t}~,~\nu=\frac{1}{2}.
\end{align}
However, by introducing a time-delay term in the equation of state parameter,
i.e. $\rho\left(  t\right)  +p\left(  t\right)  =\alpha\rho\left(  t-T\right)
^{\nu}$, the field equations becomes%
\begin{equation}
\dot{H}+\bar{\alpha}H\left(  t-T\right)  ^{2\nu}=0
\end{equation}
where now exact solution exists only when $\nu=\frac{1}{2}$, when reduce to
the linear equation (\ref{bv.09}). Consequently, $H\left(  t\right)
=H_{0}\cos\left(  \omega t+\theta\right)  ,$ that is,%
\begin{equation}
a\left(  t\right)  =a_{0}\exp\left(  \frac{H_{0}}{\omega}\sin\left(  \omega
t+\theta\right)  \right)
\end{equation}
from where it follows, that there is a periodic behaviour around the Minkowski
spacetime with radius $e^{H_{0}}$. Constants $\omega~$and $\theta$ are
determined by (\ref{bv.11}).

We conclude that the introduction of the time-delay parameter provides new
behaviours in the evolution of the cosmological model, which are of special
interest. {In the bulk viscosity model of consideration the time-delay
provides periodic behaviour when we reach or going far from a de Sitter
universe, the latter can be seen as an alternative way to escape the
exponential inflationary era.}

Moreover, {a cyclic universe around an exponential growth was proposed
recently in \cite{cl1}. Indeed, such consideration can solve various problems
of modern cosmology, from the homogeneity, isotropy and others. This work can
be seen as a mathematical description in order to achieve the a similar
universe with that proposed in \cite{cl1}. }However, in our model, the
periodic oscillation decays for specific values of the time-delay.
Specifically, the periodic is a stationary solution, however a singular
behaviour for the scale factor can be recovered. {Indeed, for large values of
the Hubble function, equation (\ref{bv.14}) becomes }$\dot{y}+\frac{3}%
{2}\gamma y^{2}\simeq0${, thus the scaling solution }$H\left(  t\right)
\simeq\frac{1}{t}${ is recovered which describes a universe dominated by an
ideal gas.}

\end{document}